\documentclass{elsart}

\usepackage{epsfig}

\begin{document}

\newcommand{\be}{\begin{equation}}
\newcommand{\ee}{\end{equation}}

\begin{frontmatter}

\title{Weak insensitivity to initial conditions at the edge 
of chaos in the logistic map\thanksref{miur}}

\author[caUni,caINFN]{M. Coraddu\corauthref{corr}},
\corauth[corr]{Corresponding author.}
\ead{massimo.coraddu@ca.infn.it}
\author[caUni,caINFMSLACS]{F. Meloni},
\author[caUni,caINFN]{G. Mezzorani},
\author[caUni,caINFMSLACS]{R. Tonelli}

\address[caUni]{Physics Dept., Univ. of Cagliari,  
I-09042 Monserrato, Italy}
\address[caINFN]{Istituto Nazionale di Fisica Nucleare,
Cagliari,  I-09042 Monserrato, Italy}
\address[caINFMSLACS]{INFM-SLACS Laboratory, Univ. of Cagliari, 
         I-09042 Monserrato, Italy}

\thanks[miur]{This 
work was partially supported by MIUR (Ministero dell'Istruzione,
del\-l'Uni\-ver\-si\-t\`a e della Ricerca) under MIUR-PRIN-2003 project 
``Theoretical Physics of the Nucleus and the Many-Body Systems''.
}

\begin{abstract}
We extend existing studies of weakly sensitive points within the framework 
of Tsallis non-extensive thermodynamics to include weakly insensitive 
points at the edge of chaos. Analyzing tangent points of the logistic map
we have verified that the generalized entropy with suitable entropic index
$q$  correctly describes the approach to the attractor.
\end{abstract}

\end{frontmatter}

\section{Introduction}
Strong sensitivity to initial conditions is one of the main features of chaos.
In the simple case of a one-dimensional dynamical variable 
$x$\/ the sensitivity function is 
$ \xi(t) = \Delta x(t) / \Delta x(0) \sim\, e^{\lambda t} \;$
for $\Delta x(0) \rightarrow 0$ and $t \rightarrow \infty$,
where a positive (negative) Liapunov exponent $\lambda$
characterizes strong sensitivity (insensitivity) to initial conditions.
The marginal case $\lambda = 0$ has already been treated 
in the framework of non-extensive 
thermodynamics~\cite{Ts:97,Ly:98,Ti:99,Ro:02,Ro:02a}, 
introduced by  Tsallis \cite{Ts:88} to describe systems with
long-range interactions or fractal space-time structures,
see Ref.~\cite{Ts:02} and references therein.
A  suitable generalization of the sensitivity function
is~\cite{Ts:97,Ly:98}:
\be
     \xi(t)\, =\, \lim_{\Delta x(0) 
\rightarrow 0 } \lim_{t \rightarrow \infty  } \, 
           \frac{\Delta x(t)}{\Delta x(0)}\, \sim\, 
   [ 1 + (1-q_{sen}) \lambda_{q_{sen}} t ]^{1/(1-q_{sen})} \; \; .
    \label{eq:sensitivity}
\ee
The parameter $q_{sen}$ is related to
the entropic index $q$, that controls the
degree of non-extensivity in the entropy 
introduced by Tsallis ($k_B = 1$):
\be 
    S_q\, =\, \frac{1\, -\, \sum p_i^q}{q\, -\, 1} \; .
    \label{eq:entropy}
\ee 
The usual exponential sensitivity, $\xi \sim e^{\lambda t}$, and 
the standard extensive entropy, $S_1 = - \sum p_i \ln p_i$,
are reproduced in the $q \rightarrow 1$\/ limit, while
an entropic index $q>1$ ($q<1$) characterizes the 
so-called weak insensitivity (sensitivity)
to initial conditions~\cite{Ts:97,Ly:98}.
The logistic map
\be
 x_{t+1} = f(x_t) = 1 - \mu x_t^2 \; \; \; ; \; \;
    -1\leq x_t \leq 1\, ,\; 0 \leq \mu \leq 2\, , \; t = 0, 1, 2, \ldots 
    \label{eq:logistic}
\ee
is a simple one-dimensional system that can be used to investigate 
the connection between dynamical 
behavior and non-extensivity at the edge of chaos.
The Liapunov exponent vanishes for specific values of the parameter $\mu$:
these points can be $2^\infty$ bifurcation  critical points,
which are weakly sensitive ($q<1$), and periodic and tangent bifurcation 
points, which are weakly insensitive ($q>1$). \\
Transition from  periodic to chaotic behavior at the 
$2^\infty$ bifurcation critical points ($\mu_c$) involves a 
Feigenbaum-scaling cascade of period doubling, 
while the opposite transition at the tangent bifurcation points ($\mu_t$) 
shows intermittency. 
In spite of the great  amount of work on the
weakly-sensitive critical points, $\mu_c$, leading to the determination
of the  entropic index using different 
approaches~\cite{Ly:98,Ti:99,Ro:02,Mo:00,La:00,Ba:02},
much less attention has been given to
weakly-insensitive points at the edge of chaos:
these latter points are the subject of our study. 
\section{Numerical analysis}
For $\mu$ that satisfies
$f^{(k)}_{\mu} (x) = x$ and $|f^{(k)'}_{\mu} (x)| = 1$,
the  Liapunov exponent $\lambda =0$ and $\xi(t)$ of 
Eq.~(\ref{eq:sensitivity}) has a power-law behavior 
with $q_{sen} > 1$ (weak insensitivity).
All the bifurcation points 
and the tangent points $\mu_t$\/ at the beginning of the periodic 
windows (at the edge of chaos) belong to this ensemble.
The value of  $q_{sen}$  can be derived 
within the continuous limit approximation~\cite{Ts:97,Ro:02a}; for instance,
$q_{sen}= 3/2$, or $(1- q_{sen})^{-1} = -2$, 
at the beginning of the period-three window ($\mu_t = 7/4$),
while $q_{sen}= 5/3$, or $(1- q_{sen})^{-1} = -3/2$, at 
the first bifurcation point of the main sequence ($\mu = 3/4$).

\begin{figure}[ht]
\begin{center}
\epsfig{figure=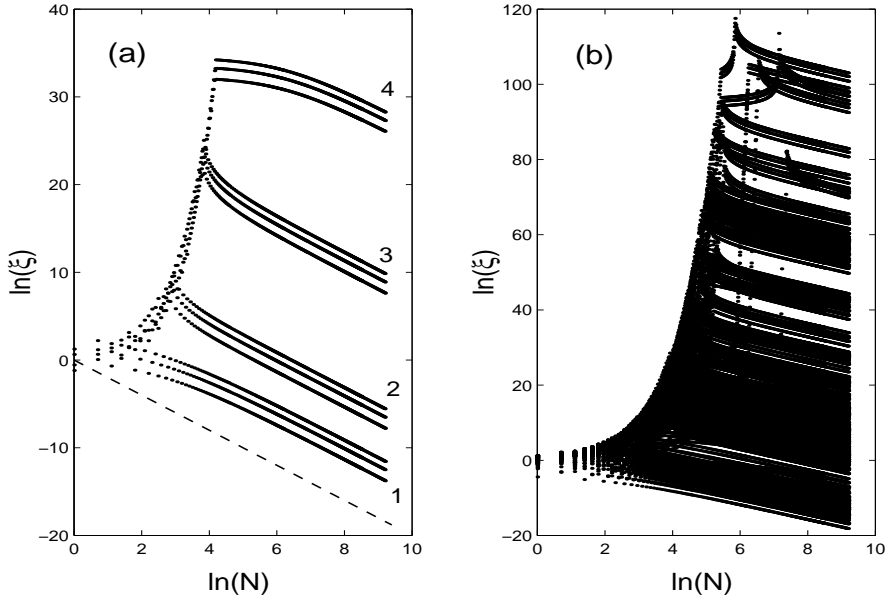,height=8cm}
\caption{$ \ln(\xi(N))$  vs $\ln(N)$ for $\mu_t =7/4$.
In frame (a) the label 1 (2, 3, 4) refers the initial condition
$x_0$=0 ($x_0$=0.788149612215968, $x_0$=0.937389375675543, 
$x_0$=0.691124729036456);
the dashed line $\xi = N^{-2}$ is a guide for the eyes. 
In frame (b) the same function is plotted for
100 randomly-chosen initial conditions.
\label{fig:SensitivFunc}
}
\end{center}
\end{figure}

Calculating the logarithm of $\xi(t)$,
approximated as $ \sum_{i=1}^{N-1} \ln(|2 \mu x_i|) $, for
the first $N \gg 1$ time steps,
we verified that the sensitivity function decreases as a power. 
In Fig.~\ref{fig:SensitivFunc} $\ln(\xi(N))$ is plotted
versus $\ln(N)$ for an ensemble of starting points $x_0$.
The system shows two regimes: (1)
at the beginning it converges to the attractor 
(the three almost-stable solutions of the 
equation $f^{(3)}_{\mu=7/4}(x) = x$); (2)
then $\xi(N)$ shows the expected 
asymptotic behavior $\sim N^{-2}$.

The initial convergence to the attractor strongly depends on the
initial condition $x_0$;  for instance frame (a) of
Fig.~\ref{fig:SensitivFunc}
shows that $\xi(N)$ decreases in the first few steps for $x_0 = 0$,
while it grows exponentially and for a longer time for other
initial conditions. In general most of the initial conditions lead
to an  initial exponential growth, see frame (b) of  
Fig.~\ref{fig:SensitivFunc}.
This result, typical of chaos, is not trivial and indicates a 
not-smooth (probably fractal) basin of convergence to the attractor;
as a counterexample there is no initial  exponential growth for
$\mu = 3/4$.\\
The same conclusion about the initial regime
is reached with a completely different 
numerical experiment: 
following Refs.~\cite{La:99,La:00},
we have partitioned the interval $[-1, 1]$ into $W=10^4$ equal cells, taken
$N_0=10^6$ random initial points  $x_{0}$ inside one of the cells
({\em concentrated initial condition}), and let them evolve according to
Eq.~(\ref{eq:logistic}); then we have
repeated the experiment 1000 times randomly changing the initial cell. 
The initial spread was so fast that more than 5000 cells ($W/2$) were
occupied after 30 time steps in 72\% of the cases,
in agreement with an exponential growth in the
initial part of the evolution. \\

\begin{figure}[ht]
\begin{center}
\epsfig{figure=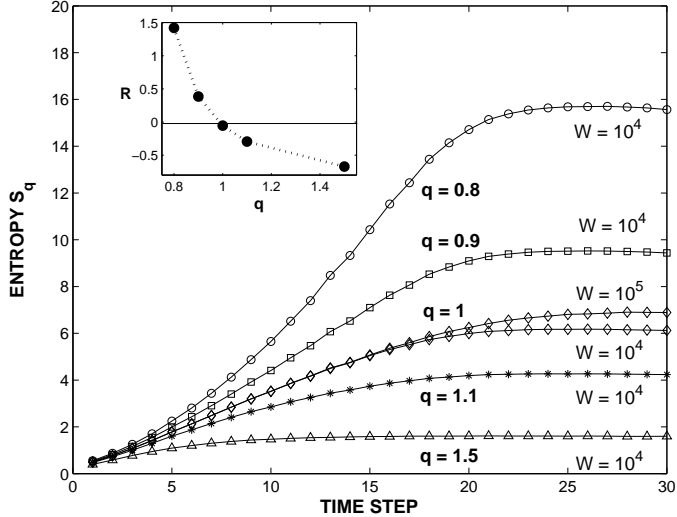,height=7cm}
\caption{Entropy evolution for several $q$'s at $\mu = 7/4$ and $N_0$
initial conditions concentrated in one cell; $S_q$ has been averaged 
over 1000 initial cells. All curves with $W=10^4$ have $N_0=10^5$,
the one curve with  $W=10^5$ ($q=1$) has $N_0=3 \times 10^5$. 
The inset shows  the non-linearity coefficient $R$ vs $q$ 
for $3=t_1< t < t_2 = 12$.
\label{fig:ChaoticRegionEntropy}
}
\end{center}
\end{figure}
In Refs.~\cite{La:99,La:00}, introducing the generalized Kolmogorov-Sinai entropy
$ K_q = \lim_{t \rightarrow \infty} 
\lim_{W \rightarrow \infty} \lim_{N \rightarrow \infty} (S_q(t) - S_q(0))/t $\/
and identifying the probability $p_i(t)$\/ by the accumulation number of the cell-i:
$p_i(t) \equiv N_i(t)/N_0$, 
it was conjectured~\cite{La:99,La:00} that $K_q$ is finite in
a given system only for a specific value of $q$, with $q=1$ corresponding
to the chaotic behavior. \\
We have extended this conjecture in two directions:
(a) we have considered the asymptotic  behavior of $S_q(t)$ for
weakly insensitive points at the edge of chaos, where entropy
decreases; (b) we studied $S_q(t)$ in the pre-asymptotic (exponential)
convergence to the attractor.

In Fig.~\ref{fig:ChaoticRegionEntropy} we show the
pre-asymptotic convergence to the attractor 
of  $S_q(t)$ averaged over the above numerical
experiments, all of  which 
had $S_q(0)=0$ (concentrated initial conditions),
for different values of $q$:
only $q=1$ is compatible with a linear growth of $S_q$.
In fact, the inset shows that 
the nonlinearity coefficient $R \simeq 0$ for $q=1$,
where $R \equiv C(t_1 + t_2)/B$, with
$A + B t + C t^2$ used to fit
$S_q(t)$ between $t_1$ and $t_2$.  
The saturation maximum value would correspond to an uniform distribution
$p_i=1/W$, {\em e.g.}, $S_{max} = \ln(W)$ for  $q=1$, and increases with
$W$.
Both the linear growth of $S_1(t)$ in Fig.~\ref{fig:ChaoticRegionEntropy}
and the exponential growth of $\xi(t)$ in
Fig.~\ref{fig:SensitivFunc} are compatible with a chaotic dynamics 
of the system in the first pre-asymptotic regime. 

Since the asymptotic regime is instead characterized by a power
behavior, as shown in Fig.~\ref{fig:SensitivFunc}, we
conjectured that $S_q(t)$ were linear and, therefore, $K_q$ finite
only for  $q \neq 1$ in these weakly insensitive points,
analogously to what happens at the weakly sensitive points.

As proposed in~\cite{Ba:02},  we selected the initial cells that give a faster spread of the
distribution: the asymptotic behavior starts from about the 
maximum value of the entropy and these cells give the
dominant contributions. 
\begin{figure}[ht]
\begin{center}
\epsfig{figure=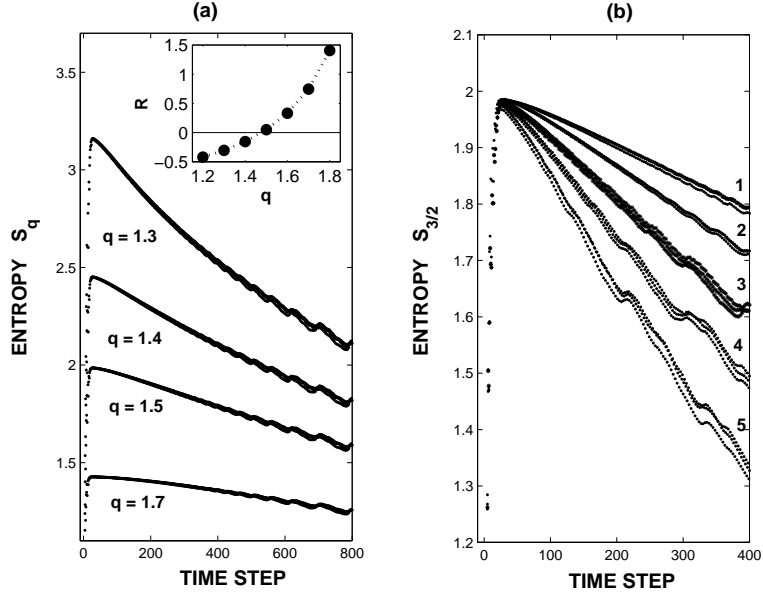,height=8cm}
\caption{
Entropy evolution at $\mu = 7/4$ with
initial conditions concentrated in one cell: 
(a) for $W$=512000 and   $q=1.3$, 1.4, 1.5 and 1.7;
(b) for $q=3/2$ and $W=512000$\/ (256000, 128000, 64000, 32000)
for the ensamble of points 1 (2, 3, 4, 5).
Curves 2-5 in (b)
have been averaged on five choices of the initial cell, 
curve 1 has not been averaged.
The inset in (a) shows  the non-linearity coefficient $R$ vs $q$ 
for $51=t_1< t < t_2 = 550$.
\label{fig:AttractorRegionEntropy}
}
\end{center}
\end{figure}
Fig.~\ref{fig:AttractorRegionEntropy} shows $S_q$ 
for different values of $q$ and $W$.
$S_{max}$ is approximately reached in the
first $\sim$ 30 steps, then $S_q$ decreases. 
This implies an overshooting in the $S_q$\/ time evolution.
In particular,
frame (a) demonstrates that only when $q = q_{sen} = 3/2$
$S_q$ decreases linearly (and $K_q$ is finite): precisely
the value expected~\cite{Ts:97,Ro:02a} from the 
attractor structure at $\mu_t = 7/4$ and characterizing
the long-time  behavior of $\xi(t)$ in Fig.~\ref{fig:SensitivFunc}.
Confirming our hypothesis,
the conjecture on the generalized Kolmogorov-Sinai entropy 
retains its validity also for weakly sensitive 
conditions at the edge of chaos:
a finite $K_q < 0$ is found only for $q = q_{sen} > 1$. 
Frame (b) in fig.~\ref{fig:AttractorRegionEntropy}  shows that
the slope of the linear part of $S_{3/2}$ depends on
the number of cells $W$: the larger $W$, the longer the
linear part of $S_{q}(t)$ and the more time is needed to
reach the asymptotic value.
\section{Conclusions}
We have studied the sensitivity to initial condition $\xi(t)$
and the entropy at a
weakly insensitive point at the  edge of chaos 
($\mu=7/4$) finding two regimes.

During the initial pre-asymptotic regime the dynamics mimics a
chaotic behavior: the sensitivity to the initial conditions
is exponential, the Shannon entropy  grows linearly with time, and
the Kolmogorov-Sinai entropy is finite as shown 
by Figs.~\ref{fig:SensitivFunc} and \ref{fig:ChaoticRegionEntropy}.

In the subsequent asymptotic regime $\xi(t)$ decreases with a
power law characterized by the  index $q=q_{sen}=3/2$,
see Eq.~(\ref{eq:sensitivity}) and Fig.~\ref{fig:SensitivFunc}.
The generalized non-extensive $q$-entropy of Eq.~(\ref{eq:entropy})
has a linear (decreasing) behavior only for $q=q_{sen}=3/2$, and
for the same $q$ the generalized Kolmogorov-Sinai entropy $K_q$
is finite.

These results suggest that the conjecture on the behavior
of entropy and on the value of the entropic index $q$
for weakly sensitive points can be extended also to 
weakly insensitive points.

We are completing a study of the same problem with 
the so-called relaxation based approach~\cite{Mo:00},
which appears to corroborate our conclusions.

\end{document}